# Whole-genome modeling accurately predicts quantitative traits in plants.

Laurent Gentzbittel[1,2,*], Cécile Ben[1,2], Mélanie Mazurier[1,2], Min-Gyoung Shin[3], Martin Triska[4], Martina Rickauer[1,2], Yuri Nikolsky[5,6], Paul Marjoram[7], Sergey Nuzhdin[3], Tatiana V. Tatarinova[4,8]

1. Université Fédérale de Toulouse ; INP ; EcoLab (Laboratoire Écologie Fonctionnelle et Environnement), ENSAT, 18 chemin de Borde Rouge, 31326 Castanet-Tolosan, France
2. CNRS-EcoLab (Laboratoire Écologie Fonctionnelle et Environnement), 31326 Castanet-Tolosan, France
3. Molecular and Computational Biology Program, University of Southern California, Los Angeles, California 90089, USA
4. Spatial Sciences Institute, University of Southern California, Los Angeles, California 90089, USA
5. School of Systems Biology, George Mason University, Manassas, Virginia 20110, USA
6. Vavilov Institute of General Genetics, Moscow, Russian Federation
7. Division of Biostatistics, Department of Preventive Medicine, Keck School of Medicine, University of Southern California, Los Angeles, California 90089, USA
8. Department of Pediatrics, Keck School of Medicine and Children's Hospital Los Angeles, University of Southern California, Los Angeles, California 90027, USA
* E-mail: gentz@ensat.fr

## Abstract.

Understanding the relationship between genomic variation and variation in phenotypes for quantitative traits such as physiology, yield, fitness or behavior, will provide important insights for both predicting adaptive evolution and for breeding schemes. A particular question is whether the genetic variation that influences quantitative phenotypes is typically the result of one or two mutations of large effect, or multiple mutations of small effect. In this paper we explore this issue using the wild model legume Medicago truncatula. We show that phenotypes, such as quantitative disease resistance, can be well-predicted using genome-wide patterns of admixture, from which it follows that there must be many mutations of small effect.

Our findings prove the potential of our novel "whole-genome modeling" –WhoGEM– method and experimentally validate, for the first time, the infinitesimal model as a mechanism for adaptation of quantitative phenotypes in plants. This insight can accelerate breeding and biomedicine research programs.

# Introduction.

All living organisms adapt to the changing environment. The adaptive traits of humans include, among others, skin pigmentation [1], altitude adaptation [2] and lactose tolerance [3]. Plants may alter flowering time, length of developmental stages, or photosynthesis [4]. Selective pressure is often imposed by geographic variables such as climate conditions, by pathogen exposure or food resources. Thus, in free living species, genetics and geography are closely and measurably associated [5]. The species respond by changing population structure via migration, by allele sorting due to random events (genetic drift) and by natural selection [6]. Local adaptations among populations are a response to heterogeneity of the natural landscape. In general, a locally adapted population has higher fitness at its native site than any other population introduced to that site [7–9]. Typically, local adaptation favors biodiversity, as the alleles are selected under distinct environmental conditions [10]. In most cases adaptive traits represent measurable phenotypes (i.e. "quantitative traits"), such as height, yield or pathogen resistance, that depend on the cumulative actions of many genes, with variants occuring across multiple loci with often poorly understood relationships between them [7].

When adaptation affects one or a few loci with large individual effect, its inheritance is explained by classical Mendelian and population genetics. The set of variants leading to adaptation is rapidly fixed in the population, along with neighbouring genomic regions, according to the "selective sweep" model [11]. Selective sweeps create a genomic signature which consists of reduced genetic diversity and extended linkage disequilibrium in the genomic region surrounding the loci under selection [12–14]. In plants, major selective drivers are associated with selective sweeps, for example for soil conditions [15], and climate adaptation [16–18]. For most candidate signals, we know neither the exact effect on phenotype, nor the nature of the selective pressure [12]. This lack of understanding of the functional mechanisms is the major drawback of the "selective sweep" model [19]. Due to the existence of multiple mechanisms (parallel adaptive pathways, redundant signaling and metabolic networks, etc.), an adaptation to the same selective stress can be achieved by different combinations of sweeps [20], resulting in a large heterogeneity of "sweeps" across populations. Therefore the signature's content is highly dependent upon population structure and restricted gene flow, particularly in plants [21].

When adaptation has a polygenic basis, the quantitative trait will evolve rapidly, via small changes in the population frequencies of a large number of pre-existing polymorphisms [22]. Adaptation of quantitative traits is described by the "infinitesimal model", proposed by Fisher [23]. This approach originally consists of purely phenotype-based modeling, with no regard to molecular basis of inheritance. Consequently, in classical animal and plant breeding, the degree of expression of selective traits (size, yield etc...) are predicted based on the performances of the individuals and their relatives, ignoring molecular markers. Such phenotypic prediction makes the models very robust [24], and they are widely applied in breeding and in some evolutionary and ecological studies [25;26]. However, phenotype-based models obviously lack the biological mechanisms behind the phenotypic traits.

Other quantitative methodologies link molecular data to phenotype via Quantitative Trait Loci (QTLs). One set of methods is focused on studying the QTLs within relatively small families (typical for human diseases) [27] and in the context of controlled crosses in cultivated plants [28]. The other approaches deal with large populations, focusing on unrelated individuals, such as Genome-Wide Association Studies (GWAS) [29;30]. In either case, the studies follow a filtering workflow aimed at identifying relatively few QTLs or variants that are most relevant to the phenotype (be it a trait or a disease). Over recent years, this approach, fueled by increasingly affordable genome sequencing or genotyping, has led to an explosion of disease-related gene discoveries in human [31] and has become a method of choice in plant and animal breeding [32] and in studies of adaptations in natural populations [29].

However, GWAS (and QTL-based methodology in general) have also substantial drawbacks. First, only a few (if any) SNPs or QTLs are likely to have statistical significance in any given GWAS on human or plants [33]. Therefore, the variants found by GWAS typically explain only a minor fraction of the heritability of a specific trait (10-30%), making them poor predictors of phenotypes [34]. To address the weak predictive power of GWAS-derived variants for phenotype (so-called "missing" heritability), some scientists have adapted the whole-genome prediction method proposed by Meuwissen et al. [35]. Such methods simultaneously use the full set of genome-wide SNPs to predict phenotypes [36]. It is, however, difficult to correctly allow for population structure within these methods. Second, due to high level of genetic heterogeneity and epigenetic modification in organisms, most SNPs and QTLs are "unstable", i.e their statistical relevance to phenotype is not supported by follow-up studies [37]. These drawbacks substantially limit applicability of the results of QTL or "big data" omics studies in clinical settings (such as disease predisposition multi-gene panels) or in plant and animal breeding [38].

Here, we offer a novel empirical paradigm, that we call "whole-genome modeling" -WhoGEM – for testing polygenic adaptation. Essentially, we consider the complete, genome-wide universe of genetic variability, spread across several ancestral populations originally separated (e.g. by climate, geography, tissue differentiation), ultimately displaying distinct phenotypes. Assuming that each individual is a descendant of one or several ancestral populations, the relative contributions of such populations to the genome of each individual – the so-called admixture components – are estimated using a likelihood algorithm. Our innovative working hypothesis postulates that a large proportion of current phenotypic variation between individuals may be best explained by population admixture. We therefore use admixture proportions, instead of SNP-based analysis of genome scans, to test for correlation between genomic variation and quantitative phenotypes, or environmental variables. As such, our method combines phenotypic anchoring of molecular markers with "whole-genome modeling" of the genotype, avoiding the major drawbacks of the abovementioned models by explicitly integrating the population structure.

Unlike animals, plants feature complex mating systems including selfing and limited gene dispersal through seeds and pollen, and a distinct immune system. Importantly, plants must survive under permanent selective pressure from local environmental conditions. These features make plants excellent objects for testing polygenic adaptation hypotheses [8]. We test our paradigm on Medicago truncatula, a model legume with detailed genomic data available across a number of the Mediterranean populations [39–42]. As a short-lived and self-compatible species, M. truncatula probably has a more differentiated population structure and is expected to be a better model to study local adaptation than long-lived or outcrossing species [43]. Existing knowledge of the exact population structure of M. truncatula was incomplete, with contradictory versions described [44;45]. As this is one prerequisite for our study, we thus infer that structure with precision.

We define admixture components using a high-quality SNP set, based on the genome information of 262 M. truncatula accessions located around the Mediterranean Basin [39]. As a calculation tool, we adapt the recent Geographic Population Structure algorithm [46] to plants (plantGPS). We use plantGPS to refine our understanding of population structure and assess the optimal number of admixture components. We demonstrate that WhoGEM, based on an admixture model, accurately predicts quantitative traits. Focusing on disease resistance, we predict resistance to root pathogens. Using an independent set of accessions not previously characterized, we experimentally validate the prediction of disease resistance levels from whole-genome data. We also unveil relations with admixture components by analysing other key quantitative functional traits, such as plant height. Our analysis shows that geographic adaptations also rely upon a polygenic basis. Our results are among the first to experimentally validate Fisher's infinitesimal model for the adaptation of a quantitative phenotype in plants. We also argue that our WhoGEM method is directly applicable to a wide range of biomedicine problems, such as prediction of drug response and carcinogenesis, and to accelerate breeding programs in agriculturally important plants and animals.

## Results.

### A pattern of eight discrete populations around the Mediterranean Basin is revealed by admixture-based analysis of the Medicago truncatula genome.

Many population genetic analyses assume a unique origin for the populations, with stepping stone divergence patterns. We take a different approach, relying on the glacial refugia hypothesis postulated for M. truncatula [47], and nterpret our data assuming refuge areas with subsequent plant population growth and admixture. To define M. truncatula populations, we conduct a three-step analysis combining admixture-based tools, Principal Component Analysis (PCA) and plantGPS.

First, using a likelihood-based admixture analysis [48], we identify a substructure at $K = 7$ and $K = 8$ in which individuals appear homogeneous in their admixture composition (Figure 1, Supplementary Figure 1). Higher K values yield noise which appears as ancestry shared by very few individuals within the same populations.

Second, the most suitable number of admixture components is verified using a PCA-based analysis. For $K = 9$ and higher, putative populations are not independently spread in the PC space, adding a supplementary argument for $K = 8$ being the optimal maximum number of source populations (Supplementary Figures 2 to 5).

Third, we apply our novel tool plantGPS to evaluate the accuracy of geographic assignment for each sample, using the distance between its recorded and predicted location. The plantGPS tool uses the admixture components of each sample. The empirical cumulative curve of distances between predicted and observed locations shows that $K = 8$ results in best predictions (Supplementary Figure 6). With eight admixture components, 50% of accessions have their location predicted to within 100 km of their recorded location, and 75% within 800 km.

This three-step analysis demonstrates that population structure in M. truncatula can be adequately explained using eight admixture components. Pair-wise Wright's $F_{ST}$ divergences [49] between the admixture components (comparing the variance in allele frequencies among the components) indicate that they are strongly differentiated (Table 1a). Therefore, we use this number of components for subsequent analyses. The structure of the current M. truncatula population is obtained by hierarchical clustering of the samples, based on their admixture patterns (Supplementary Figure 7, Supplementary Table 2).

Figure 2 displays the distribution of the eight putative M. truncatula populations around the Mediterranean Basin, showing the genome admixture proportions of the 262 samples. Based on this picture, we assign each population to a representative geographical region (Table 1b). Estimates of $F_{IS}$ values (the inbreeding coefficient of an individual relative to its sub-population) are similar among the eight populations, suggesting no obvious intra-population heterogeneity (Table 1b). All populations are clearly differentiated, even over short geographical distances, such as with the two Spanish populations.

Relationships among the populations are estimated based on the 840K SNP dataset, using accessions that have at least 90% of their genome assigned to a given ancestral population to represent that population. The resulting dendrogram (Supplementary Figure 8) reveals two main clades that reflect the major divergence event. Clade 1 contains populations from the South West of the Mediterranean Basin: "Algiers" (K1), "Spanish Coastal" (K2) and "Spanish Morocco Inland" (K8). Clade 2 contains the accessions from the North East of the Mediterranean Basin. We speculate that the clade 1 / clade 2 divergence reflects expansion from glacial refugia during the early Holocene[50]. Within clade 2, the "French" (K6) population is clearly separated from the "Greek" (K7) population, in agreement with the "Maritim and Ligurian Alps" glacial refugia hypothesis [50].

## Geographical and bioclimatic variables significantly shape part of genetic variation in M. truncatula.

The aspects of environmental variation that generate selective gradients are poorly understood for most species. The repartition of the eight M. truncatula populations in the climatic zones defined by the Koppen-Geiger climate classification [51] shows that several populations are present in the same climate zone; and that the "Greek" population, is scattered through several climatic zones (Supplementary Figure 9). This makes it unlikely that global climate types shaped populations.

We check for associations between admixture components and 19 local bioclimatic variables, defined by WorldClim (http://www.worldclim.org). Correlations between the proportions of the various admixture components and any of the bioclimatic variables are shown in Supplementary Table 3, along with tests of significance (Supplementary Table 4). Associations with bioclimatic variables depend strongly on the admixture component. For example, the "Spanish Coastal" component is negatively correlated with temperature seasonality (BIO4) and temperature annual range (BIO7), indicating that this genome is present in accessions that grow in regions with moderate annual temperature and small temperature seasonal contrasts. Interestingly, the admixture proportions of the "North Tunisian Coastal" population are not related to any bioclimatic variables, suggesting that the differentiation of this genome may be due to other factors. Friesen et al. [52] described how accessions belonging to this population harbor alleles that assort non-randomly with soil salinity, suggesting a differentiation of the "North Tunisian Coastal" population due to this particular abiotic condition.

Next we use redundancy analysis (RDA [53]) to partition genomic variation within the species, summarized by the admixture proportions, into components explained by climate and geography. It allows estimating the change in the structure of genomic variation across spatial scales (latitude, longitude and elevation) and climatic variables. Figure 3 shows that about half of the genomic variation is due to climate or geography ($r^2 = 0.46$; $P \leq 0.001$), with climate as a major source of variation in admixture component (41.4%). The variation explained jointly by geography and climate is 17.6%, geography alone contributes to 5% only.

This partition of genomic variation in response to climate is clearly different to A. thaliana, in which Isolation By Distance (IBD) is important and climate variation among sites of origin explained only slightly more genomic variation than geographical distance [54]. Here we show that genome admixture proportions are correlated to bioclimatic variables. This relationships suggests that a large number of loci/genes are involved in response to climate, even if the phenotypes involved in this response are still unknown.

## Genome admixture components predict quantitative resistance to plant pathogens, as demonstrated by experimental validation.

Altogether, the above results suggest that a significant part of the variation in genome admixture components is due to other factors, such as soil or biotic interactions. Consequently, we explore whether M. truncatula population structure might relate to adaptation for polygenic traits such as plant-microbe interactions.

Two types of disease resistance are described in plants (i) complete resistance conditioned by a single gene[55] and (ii) partial resistance, also called quantitative disease resistance (QDR), conditioned by multiple genes of partial effect[56]. QDR often confers broad-spectrum resistance, being predicted to be critical for efficient control of epidemics. It is characterized by a continuous range of phenotypes from susceptible to fully resistant. QDR is often described by QTL that support the resistant phenotype and suggest modes of polygenic adaptation[56]. Studies that attempt to dissect a QDR trait have reported genes with various biological functions such as ABC transporters[57] or atypical kinases[58]. However, these genes do not explain all of the genetic variances reported using controlled crosses or GWAS studies. The aggregating of QDR loci has been useful in decreasing disease symptoms[59], but fully resistant phenotypes are rarely described, suggesting that additional, possibly numerous loci, are required.

M. truncatula is prone to infection by the soil-borne fungal root pathogen Verticillium alfalfae. Verticillium wilt response in M. truncatula is a QDR, regulated by QTLs that differ across resistant accessions and vary according to the fungus strains[60;61]. Both plant and fungal species co-exist around the Mediterranean Basin (CABI database, PlantWise database http://www.plantwise.org/, August,25 2015).

Figure 4 depicts the geographical partition of the Maximum Symptom Score (MSS) of 242 M. truncatula accessions when infected with the V. alfalfae strain V31-2 (Supplementary Table 11), together with their admixture patterns. Accessions located west of the Mediterranean Basin are mainly resistant to the V31-2 strain (low MSS), while accessions located east of the Mediterranean Basin are susceptible. Testing WhoGEM's working hypothesis, we explore whether the degree of quantitative resistance to V. alfalfae can be explained by values of admixture components. Our findings (Table 2) show that the proportions of four admixture components are significantly related to MSS, as suggested by Figure 4 ($r^2 = 0.31$, $P \leq 2.2 \times 10^{-16}$). The average MSS value of the "Spanish Coastal", "Spanish-Morocco Inland" and "South Tunisian Coastal" genome components are 1.04, 1.6 and 1.71 respectively, making them resistant genomic backgrounds. The average MSS value of the "Greek" genome component is ˙3, making it a clearly susceptible genomic background.

We experimentally validate these computational results using an independent set of accessions not re-sequenced previously. We predict the phenotypes from admixture components using the model resulting from our above analysis as follows: uncharacterized accessions located within the geographic zone of resistant (respectively susceptible) accessions should exhibit resistant (respectively susceptible) phenotypes. We test 32 new accessions from the "Spanish Coastal" or "Spanish-Morocco" geographic zone, and 39 new accessions from the "Greek" geographic zone (Figure 5a), and assess their disease resistance level (Supplementary Table 12).

Figure 5b shows the observed MSS of accessions predicted to be resistant or susceptible, along with MSS of Spanish-Moroccoan (resistant) and "Greek" (susceptible) reference samples. MSS values of samples expected to be resistant or susceptible differ significantly (ANOVA $P < 2 \times 10^{-16}$) and samples predicted to be resistant are significantly different from those predicted to be susceptible (adjusted $P < 2 \times 10^{-16}$). Interestingly, multiple means comparisons show that predicted resistant (or susceptible) accessions are not significantly different from their respective reference populations (adjusted $P = 0.55$ and $0.90$ respectively, Table 3). Defining the value MSS= 2 as a threshold for resistant (MSS < 2) or susceptible (MSS ≥2) accessions, we find that 26/32 of the samples predicted to be resistant from their admixture proportions actually are resistant, while 27/39 samples predicted to be susceptible are actually susceptible ($\chi^2 = 16.03$, $P = 6.22 \times 10^{-5}$). This shows that patterns of QDR can be predicted from genomic admixture analysis, arguing for the utility of WhoGEM when predicting complex phenotypes.

We have thus shown experimentally validated theoretical predictions, by WhoGEM, of the QDR level in M. truncatula. We show that the phenotypic difference between predicted resistant and susceptible accessions is around two points on a scale from 0 to 4, i.e. 50% of the phenotypic difference between the extremes of the phenotype distribution. This value is far greater than those previously reported (0.28-0.5) for phenotypic differences between alleles at the major QTLs detected in response to V. alfalfae [60;61]. Given the estimated narrow sense heritability of the trait [60;61], we suggest that genome admixture components are explaining the majority of the genetic control of this disease.

QDR is typically broad-spectrum, making the arms race between hosts and pathogens probably not critical and, consequently, currently not reported in the literature. Our results re-enforce the idea that QDR in plants is likely to result from changes to a large number of genes scattered throughout the genome, and that this is reflected in admixture proportions. Because of the co-occurrence of both the plant species and the pathogen around the Mediterranean Basin, we hypothesize that the observed pattern of quantitative resistance in the M. truncatula/V. alfalfae pathosystem may be due to natural selection, with unknown additional contributions from drift and migration.

### Genome admixture components can be significant predictors of quantitative functional traits in plants.

Knowledge of the selective pressure acting on the phenotype, could lead to insights into the respective contributions of adaptive selection and drift toward phenotypic differentiation among populations. This can easily be tested using plant-microbe interactions, in particular for QDR, by comparing the geographical distribution of plants and pathogens.

Aphanomyces euteiches is a soil-borne pathogen of legume crops, mainly occurring North of the 45th parallel. Using data reported by Bonhomme et al.[45], we depict the geographical structure of root-rot index (RRI), a typical quantitative phenotype measuring susceptibility of M. truncatula to this oomycete, together with the admixture patterns of the studied accessions (Supplementary Figure 10). We test whether the proportions of admixture components (Supplementary Table 2) are predictors for RRI and find a significant relationship ($P = 1.8 \times 10^{-7}$, Supplementary Table 5). The WhoGEM model accounts for 19.2% of variation in the phenotype, and provides a lower bound for heritability. Intriguingly, the populations of the Maghreb area show a contrasting response to the pathogen, whose presence is not reported in that geographic zone (CABI database, PlantWise database http://www.plantwise.org/, Aug.,25 2015; Bonhomme et al.[45]). Hence, we hypothesize that the phenotypic differentiation among the resistant and susceptible populations may be due to genetic drift or migration. This hypothesis also suggests that the cost of resistance may be negligible in the absence of pathogen, in contrast with previous results described for foliar pathogens [62]. An alternative hypothesis is that resistance to A. euteiches is driven by, or linked to, resistance to other factors, as suggested by Djebali et al. [63], and as such, not a consequence of natural selection acting toward resistance to the oomycete.

Having demonstrated that genome admixture proportions can be used to predict different QDR, we further test our approach by examining relationships between admixture component proportions (Supplementary Table 2) and several quantitative functional traits related to development as reported by Stanton-Geddes et al. [64]. Supplementary Figures 11d & e depict the geographical structure of plant height and leaf number combined with admixture proportions of the recorded accessions. Plant height (Supplementary Table 6a) and the number of leaves (Supplementary Table 6b) exhibit different results regarding association with genome admixture components. The influence of population structure on plant height is very significant ($r^2 = 0.21$, $P = 2 \times 10^{-11}$), but less so on number of leaves ($r^2 = 0.05$, $P = 7 \times 10^{-4}$). The results suggest that a latitudinal cline for leaf numbers may exist, with accessions south of the Mediterranean Basin harboring more leaves.

## Long-term positive selection is weakly involved in current population structure.

The above results suggest that response to selective pressure contributed to the differentiation among M. truncatula populations. When trying to understand the role of natural selection, protein-coding sequences offer the great advantage in that they allow us to distinguish synonymous and non-synonymous substitutions.

The non-synonymous/synonymous rate ratio, $\omega = dN / dS$, measures selective pressure at the protein level[6]. A non-synonymous rate which is significantly higher than the synonymous rate, resulting in $\omega > 1$, provides evidence for positive protein selection, whereas $\omega < 1$ evidence purifying selection. To test if adaptation to biotic and abiotic conditions involves selection over a long period of time, as compared to rapid changes in allelic frequencies, we compute $\omega$ in each M. truncatula population.

We conduct a population-wise binomial test to determine if the dN/dS value for each gene differs from the genome-wide average dN/dS value. In each population, we obtain clear evidence of a trend towards extreme dN/dS values; the QQ-plot of p-values is shown in Figure 6. We identify 15 genes with significant purifying selection in at least one population (Table 4 and Supplemental Table 7). The majority of these are related to development (cell wall and growth) and energy (photosynthesis and correlative UV protection), both involved in establishing fitness level.

Among the eight populations, we identify only 10 genes with both $dN/dS \geq 5$ and a coefficient of variation $\geq 0.40$ (Supplemental Table 8 and Supplemental Table 9). Five of the 10 genes are transcription factors of unknown function, indicative of the importance of regulatory pathways in adaptation. The FRIGIDA ortholog exhibits a strong differential selection pressure among the eight populations, in accordance with the necessary adaptation of vernalization and flowering times over contrasted environments[65].

Interestingly, we identify 137 genes that have $\omega \geq 7$ and a low coefficient of variation (Figure 7). These genes could be involved in the overall selection response for all accessions of M. truncatula. GO term enrichment analysis of these genes does not identify specific biological processes, suggesting that all biological processes contribute to adaptation at the same pace.

Tests based on synonymous/non-synonymous comparisons are expected to be less sensitive to demographic assumptions than site frequency spectrum or LD-based methods [21]. However, they cannot distinguish between past and current selection. Our results reveal strong purifying selection acting on genes directly involved in fitness, and show that putative positive selection is affecting all biological processes. Analysis of patterns of nucleotide variation shows that observed population structure is probably due to ongoing selection rather than long-term selection. We also show that the number of genes under putative long-term selection is low.

Genome admixture component prediction supports the hypothesis that quantitative traits (forming the majority of eukaryotes' functional traits) are affected by many genes, each being under moderate to weak selection pressure. This will be true even if the overall selection pressure is high, and especially if an additive mode of action is predominant. As a consequence, the number of genes that can be detected to be under strong positive selection may be low. This agrees with our analysis of the dN/dS ratio among the eight M. truncatula populations. Our findings challenge previously reported results [39;66], that were based on a preliminary dataset including fewer samples, no population structure, and using SNPs called on a previous version of the genome.

### Geographic localization of the reference genome of M. truncatula: plantGPS confirms that genetics helps in predicting geography.

As a supplementary tool in defining the population structure, we modified the algorithm described by Elhaik et al. [46] to adapt it to plants (plantGPS), and to reflect the expected reduced levels of migrations and outcrossing in self-pollinating plants like A. thaliana and M. truncatula.

Using the plantGPS method, we infer the geographic source for the M. truncatula A17 reference genome [40]. This accession has been isolated from the Australian Jemalong cultivar (T. Huguet, personal communication) but Jemalong's origin in the Mediterranean Basin is not documented in the literature. With plantGPS, we determine a likely primary geographical position of the Jemalong-A17 accession within the "Spanish Coastal" population, as illustrated in Figure 8. The location of A17 within this population is in accordance with its resistant phenotype in response to V. alfalfae [60]. This independent result provides additional evidence of the role of admixture components in determining phenotypes.

Use of the plantGPS algorithm also provides us with an objective function to minimize i.e. the distance between predicted and reported locations, when looking for the most likely number of admixture components. This investigation shows the potential of the plantGPS method in locating unknown plant samples based on their admixture components, it may have similar applications in forensic sciences and technologies.

## Discussion.

Understanding the relationships between genotype and phenotype is a fundamental challenge in modern biology. Linking specific genomic variations with selective traits in plants and animals (yield, fitness, etc.) and human (disease predisposition, drug response, etc.) is key for many fields, from plant and animal breeding to individualized healthcare and drug discovery.

Most adaptive events in natural populations, or selected traits in breeds of domesticated species, occur via the evolution of quantitative, polygenic traits rather than via the fixation of a single (or few) beneficial mutations[22], with some exceptions, such as monogenic human diseases and a few plant traits described by Mendelian laws. Therefore, the phenotypic variability found in natural populations is due to a complex underlying genetic interplay of multiple, often unknown, loci with allelic effects affected by the environmental conditions [24;28]. Not surprisingly, the models attempting to describe genotype / phenotype relationships suffer from major drawbacks and are reductionist (focusing on relatively few genetic features). For instance, there are typically no clear "selective sweep" signals for most traits, as a result of low specificity for the corresponding model. Likewise, the "major QTL" model, which is currently the preferred concept in GWAS and controlled cross studies, shows limited performance in identifying the loci critical for breeding programs.

Here, we offer a novel methodology for predicting quantitative traits based on a whole-genome model of genomic variants and population admixture-based algorithms. We propose to call our approach "whole-genome modeling" – WhoGEM –. We move away from focusing on "large impact" variants and instead we propose calculating a simple descriptor of "mixing proportions" in individuals believed to originate from distinct ancestral populations. Our approach is akin to the calculation of phenotypic resemblance as in the whole-genome genetic resemblance, popularized by Meuwissen and others [35;67]. Unlike the latter approach, we explicitly embed the inferred population structure in our calculations, thus expanding the method's applicability. We demonstrate the relationship between admixture proportions and quantitative traits in the model legume M. truncatula and support the view that phenotype can be explained by an additive action of a large number of loci. These components of inheritance represent combinations of genes (either protein-encoding or regulatory, such as non-coding RNAs) manifested as alleles, polymorphisms and some other genetic variants. Using examples of pathogen resistance and plant development traits, we demonstrate that the admixture proportions descriptor can predict the degree of local adaptation better than the feature selection-based methods. The WhoGEM concept is likely to be expandable to all quantitative functional traits that involve complex genetic determinism. Admixture proportions offer a meta-view of genome structure, providing information integrated across the entire genome. As admixture proportions embed information from the linked loci, they may be a straightforward way to address the long-standing debate about the relative contribution of protein-coding changes (micro-evolution) versus regulatory changes (expected to act through regulatory pathways, and thus phenotypic plasticity).

To conclude, we show that genome admixture components provide strong insights into the genetic landscape of polygenic adaptation. To the best of our knowldege, this is the first analysis to experimentally validate the infinitesimal model for natural adaptation of a quantitative phenotype in plants. Comparative analysis of response to two different pathogens clearly demonstrates that phenotypic differentiation among populations, supported by an infinitesimal model, may or may not result from natural selection. We also suggest that response to climate could be investigated using an infinitesimal model. Predicting phenotypes on the basis of genome components will help in inferring future trends of local adaptation related to global climate change. Both plant and animal breeding methodologies may beneficiate from WhoGEM. Finally, we postulate that prediction of complex traits in humans, for example drug response in clinical trials or disease predisposition models, may benefit from the same general methodology. Adaptations of our method may well be applicable to "omics" data-based modeling of metastasis, clonal selection and genetic heterogeneity in cancer research. Moreover, an extension of WhoGEM would be capable of integrating and calculating admixture proportions from multiple types of genome-wide "big data", such as whole-genome genetic landscapes, epigenetics and expression profiling. Our findings contribute to the establishment of a strong theoretical and experimental corpus of methods to detect and explain signals of polygenic adaptation within genomes.

## Materials and Methods.

**SNP selection.** A set of 262 genuine Medicago truncatula accessions [39] was used to extract SNPs (http://medicago.hapmap.org). After quality checking and LD-pruning using PLINK, using the options --geno 0.05 --maf 0.01 --indep 300 60 1.3, we selected a total of 843 171 SNPs covering the eight chromosomes of the M. truncatula genome (Supplemental Table 1).

**Development of the plantGPS algorithm.** plantGPS is an adaptation of the admixture-based Geographic Population Structure (GPS) algorithm [46] to plant species. This modification takes into account ties encountered when the genetic distances between different closely related accessions are identical. Ties are also considered when computing the contribution of other reference accessions to the sample's genetic make-up. plantGPS calculated the Euclidean distance between the sample's admixture proportions and a reference dataset, for predicting the provenance of sequenced accessions. The matrix of admixture proportions was calculated with the ADMIXTURE software package [48]. The 'shortest distance' measure, representing the test sample's deviation from its nearest reference population, was subsequently converted into geographical distance using the linear relationship observed between genetic and geographic distances. The final position of the sample on the map was calculated by a linear combination of vectors, with the origin at the geographic center of the best matching population weighted by the distances to 10 nearest reference populations and further scaled to fit on a circle with a radius proportional to the geographical distance. If the smallest distance ($\Delta_{GEN}^{min}$) that represented the sample's deviation from the best matching accession was identical for several accessions, those were considered as a set of accessions. Numerical values may thus contain ties and the initial geographical position of an unknown accession was defined as the centroid of the geographical positions of the identical, or nearest accessions. The contribution of other reference accessions m = 2..N to the sample's genetic make-up might also contain ties. The computation of the weight $w = \frac{\Delta_{GEN}^{min}}{\Delta_{GEN}(m)}$ was thus modified accordingly.

To estimate the assignment accuracy of plantGPS, we used the 'leave-one-out' approach at the individual level. In brief, we excluded each reference individual from the data set, recalculated the mean admixture proportions of its reference population, predicted its biogeography, computed the geographical distance between predicted and reported locations, tested whether it is within the geographic regions of the reported origin and then computed the mean accuracy per population. More specifically, we index our individual as the j$^{th}$ sample from the i$^{th}$ population that consists of n$_i$ individuals. For all populations, excluding the individual in question, the average admixture proportion and geographical coordinates were calculated as $\bar{\theta}_m = \frac{\sum_s \theta_{m,s}}{n_m}$ where $\bar{\theta}_m$ is the parameter vector for the s$^{th}$ individual from the m$^{th}$ population, and n$_m$ is the size of the m$^{th}$ population. For the i$^{th}$ population the adjusted average will be $\bar{\theta}_i^{-j} = \frac{\sum_{l \neq j} \theta_{i,l}}{n_i - 1}$

A set of 245 genuine M. truncatula accessions with geographical coordinates (latitudes and longitudes) served as the reference set for plantGPS. Seventeen accessions, among which the Jemalong-A17 accession that is used as the reference genome [40], were of unknown origin and not included in the reference set.

**Strategy for population structure determination.** The strategy used to identify populations combines three steps : admixture-based tools, Principal Component Analysis (PCA) and plantGPS.

First, a likelihood-based analysis is run in the unsupervised mode for K = 2 to 12. We use the ADMIXTURE software package [48] based on the collection of high-quality LD-pruned SNPs. Each plant sample was characterized by a vector of n proportions that sum to one, n being the number of admixture components (i.e. n = K), possibly ranging from K = 2 to K = 12. This vector summarizes the proportion of the plant's genome that belong to each of the n admixture component. A tree that summarizes the partition of the set of accessions into successive subsets defined by their major admixture component supports K = 7 or K = 8 as providing a stable structure for population stratification (Supplementary Figure 1). Computations were conducted two times independently and came to almost identical results.

Second, the most suitable number of admixture components was verified using a PCA-based analysis. ADMIXTURE outputs the inferred allele frequencies of each SNP for each hypothetical population. This allowed us to simulate samples (called "zombies") for each hypothetical population, viewed as reconstructed hypothetical ancient-like individuals, purged of centuries or millennia of admixture. The PCA thus included both the sampled accessions and zombies, as hypothetical ancestral individuals for each hypothetical population. Supplemental Figures 2 to 5 show pairwise PCA plots of the actual M. truncatula accessions together with simulated samples (7 zombies per admixture component), for K = 7 to K = 10. For each level of K, the ancestral populations were distributed in the space of the first six eigenvectors. PCA computations were performed using the R package SNPRelate using PLINK formated files. Similar results were obtained using 30 simulated individuals per each hypothetical population, or using computations with EIGENSTRAT (data not shown).

Third, we applied the "leave-one out" cross-validation approach of the plantGPS algorithm at the "accession" level, to estimate the difference between predicted and reported location, for each sample. This procedure was repeated for K = 2 to K = 12 and the empirical cumulative distribution of the geographical distances was used as a criteria to estimate the accuracy of geographical prediction at each number of putative genome components (Supplementary Figure 6).

We then established the definitive correlation between geographic and genetic distances between pairs of individuals, for K = 8. Given the (relatively) small distances across the Mediterranean Basin, we computed a "naive" geographical distance using pairwise Euclidean distance based on the longitude/latitude reported for the accessions. The matrix of pairwise genetic distances was computed using the admixture component proportions of each accession with K = 8. The Mantel test applied to the initial matrices revealed a significant, yet moderate, correlation between geographical and genetic distances (r=0.294, $P = 1 \times 10^{-4}$). Supplementary Figure 12 shows that the linear relationship between geographical and genetic distances is restricted to distances less than 950 km. When filtering out the distance matrices for distances > 950km, the Mantel correlation coefficient raises to 0.78, a highly significant value ($P = 1 \times 10^{-4}$). We thus fitted a linear relationship between geographical and genetic distances for geographical distances less than 950 km. The regression equation is Geo = $0.204 + 4.973 \times$ Gen + with adjusted $R^2 = 0.61$ and model $P < 2.2 \times 10^{-16}$.

The average distance between original and predicted locations is 471 km, the median is 214 km and 75% of the samples are predicted to be less than 677 km from their reported origin (Supplementary Figure 13).

To provide population identification, the final admixture frequencies of the eight components for the 262 M. truncatula accessions were calculated by applying ADMIXTURE in the supervised mode. Accessions were then clustered into populations, using hierarchical clustering based on the genome admixture proportions, using Euclidean distance and the 'average' link. The name of each population name is determined by the region which is the geographical centroid of the accessions of that population.

Relationships among populations were computed based on selected individuals that had at least 90% of their genome assigned to a given ancestral population. With that aim, genetic distances were computed based on the 840K SNP dataset (R package SNPRelate) and dendrogram was computed and drawn using R packages ape and geiger.

Maps and sample locations were drawn using the rworldmap, mapplots and maptools R packages.

**dN/dS computations.** To test for selection in M. truncatula populations, ancestral alleles of truncatula clade were estimated using MrBayes [68]. MrBayes estimates ancestral alleles based on a given phylogenetic tree. We used the phylogenetic tree of Yoder et al. [69] for this purpose. In the truncatula clade, M. soleirolii, M. turbinata, and M. doliata were used as outgroups in the estimation procedure. Ancestral alleles that were predicted with probability lower than 0.9 were removed from further consideration. Non-synonymous and synonymous mutations in M. truncatula populations were then identified based on comparison to the inferred ancestral allele states. The significance of the dN/dS ratio for each gene in each population was then calculated using a two-tailed binomial test. Specifically, we assumed the dN(dS) values were distributed as Binomial(N, p), where N is the observed number of mutations in that gene and p is set to be 1/6, which is the overall ratio of $\frac{dN}{dN+dS}$ in the data. This test will then indicate genes for which the dN/dS ratio is higher or lower than is typical.

**Evaluation of quantitative resistance to Verticillium alfalfae in M. truncatula.** A set of 313 accessions of M. truncatula has been assessed for their response to Verticillium wilt, including 242 already sequenced accessions from the HapMap project [39]. M. truncatula seeds were from our own collection or obtained from the INRA Medicago truncatula Stock Center (Montpellier, France). All the M. truncatula accessions have been phenotyped using an Augmented Randomized Block Design in 3 independent replicates for the already sequenced (reference) accessions and 2 replicates for the other accessions. Between 4 and 10 plants per genotype were used in each replicate. 10-day-old plants were root inoculated as described in Ben et al. [60]. Disease development was monitored for 32 days two or three times a week and rated using a scale from 0 (no symptoms) to 4 (dead plants). At the end of the experiment, the Maximum Symptom Score (MSS) was obtained for each plant. The LS-mean of the MSS for each accession was calculated using the linear model $y_{ijk} = \mu + accession_i + block_j + \epsilon_{ijk}$ ($y_{ijk}$ the maximum disease score for the kth plant of the ith accession of the jth block ; $\epsilon_{ijk}$, the residual) using R.

**Relationship between admixture proportions and quantitative phenotypic variables.** The 19 WorldClim bioclimatic variables (30 seconds resolution, downloaded at http://www.worldclim.org/current) were extracted for each accessions' location, using the reported latitude and longitude for that accession (raster R package). The relationships between genome components, the 19 World-Clim bioclimatic variables and geography (latitude, longitude and altitude) were modeled using redundancy analysis (RDA), an approach that examines how well of the variation in one set of variables (the bioclimatic variables and/or the geography) explains the variation in another set of variables (the genome admixture proportions of each sample). The RDA was computed using the vegan R package. RDA of admixture proportions with bioclimatic variables conditionnal to geography was also computed to estimate effects of climate "corrected for" the geography.

The relationships between genome components and phenotypes were estimated using linear models. Because of dependencies among the predictors (the proportions of genome components must sum to one), a systematic search for the best minimum model was done using the leaps R package or use of the step function with both direction, employing a significance level of α = 5% as the benchmark for using a predictor.

Spatial interpolation of phenotypic traits was performed using a thin plate spline method, with a smoothing parameter of $\lambda = 0.005$, as implemented in the R package fields.

Unless otherwise stated, all computations were done using the R statistical environment[70].

Acknowledgments We thank Xavier Tassus and Christine Tayeh for providing data of plant pathogen distributions. Peter Ralph and Eran Elhaik provided valuable comments on the manuscript. We thank Jean-Marie Prosperi for providing seeds and maintaining a large part of the Medicago truncatula collections. We thank R2n for their participation in plant phenotyping. This research received no specific grant from any funding agency in the public, commercial or not-for-profit sectors. Mélanie Mazurier was supported by a PhD scholarship from the French "Ministère de la Recherche et de l'Enseignement Supérieur" and a "Visiting Student" fellowship from INPT for a stay at USC. Laurent Gentzbittel and Cécile Ben were supported by a "Visiting Scholar" fellowship from INPT for a stay at USC and US Feed the Future Innovation Lab "Climate Resilient Chickpea". Tatiana Tatarinova was supported by a "Visiting Scholar" fellowship from INPT for a stay at INPT and by NSF Division of Environmental Biology award # 1456634 .

Author contributions L.G. and C.B. conceived the experiments, and designed and carried out data analysis. M.M, M-G S. and M.T. carried out experiments and data analysis. L.G, T.T. and P.M. co-wrote the paper. All other coauthors were involved in drafting the manuscript and provided helpful feedback for the paper.

Additional information

Supplementary Information accompanies this paper at

Competing Interests The authors declare that they have no competing financial interests.

Correspondence Correspondence and requests for materials should be addressed to L.G. (email:gentz@ensat.fr)

# Tables legends

Table 1: Putative ancestral genomes, as revealed by admixture-based analysis, and populations participating to actual levels of structure in M. truncatula.
(a) Pair-wise $F_{st}$ divergences between K = 8 admixture components.
(b) Characteristics of the eight populations defined using the K = 8 admixture components. $F_{IS}$ fixation index, number of accessions per population, name of the population and main spanned countries are indicated for each population.
(c) Pair-wise $F_{st}$ divergences between the eight populations.

Table 2: Linear model between admixture components and Maximum Symptom Scores in response to Verticillium alfalfae in a collection of 242 M. truncatula accessions.

Table 3: Mean comparisons for quantitative resistance, among groups of M. truncatula reference accessions of the two 'Spanish' populations, unknown accessions sampled in Spain, reference accessions of the 'Greek' population and unknown accessions sampled around Greece
Reference : accessions with known admixture proportions ('Spanish Coastal', 'Spanish Morocco Inland' and 'Greek' populations); Predicted : accessions with unknown admixture proportions sampled in Spain or Greece; R : quantitative resistance ; S : susceptibility

Table 4: Genes under purifying selection, as determined by dN/dS analysis in each M. truncatula population.
Genome Mt4.0 and annotation Mt4.0v2 were used for computations.

# Figures legends

Figure 1: ADMIXTURE proportions for 262 M. truncatula accessions, by increasing putative K. The x axis represents accessions sorted according to their reported country and ancestries. Each accession is represented by a vertical stacked column of colour-coded admixture proportions that reflects genetic contributions from putative ancestral populations.

Figure 2: Geographical distribution of 262 M. truncatula accessions and proposed population structure.
The stratification of the collection is obtained assuming K=8. Each sub-plot represents the extent of accessions belonging to one population. At each location, a pie chart represents the admixture proportions of the accessions's genome.

Figure 3: Venn diagram of the variation partioning for genome admixture component proportions explained by climate (left) and geography (right).
Residual is the amount of genomic variation not explained by the two explanatory variables.

Figure 4: Geographical repartition of Maximum Symptom Score (MSS) in response to Verticillium alfalfae in a collection of 242 M. truncatula accessions.
The MSS scale is displayed as a color gradient. Scale of MSS index from resistant (blue) to susceptible (red) accessions is indicated on the right.
Admixture proportions of each phenotyped accessions are summarised by pie charts.

Figure 5: Predicting and validating phenotypes based on the relationship between admixture component and partial resistance to V. alfalfae in M. truncatula.
(a) Sampling of new accessions in the geographic zone of the 'Spanish Coastal' and 'Spanish-Morocco Inland' resistant populations, and in the geographic zone of the 'Greek' susceptible populations. The MSS of the reference accessions is displayed as a color gradient (see Figure 4). (b) Violin plots of MSS for reference and sampled accessions.
Reference resistant accessions are sequenced accessions belonging to the 'Spanish Coastal' and 'Spanish-Morocco Inland' populations, reference susceptible accessions are sequenced accessions belonging to the 'Greek' population.

Figure 6: QQ-plot of p-values for binomial test of dN/dS, in each of the eight M. truncatula populations, for all anotated genes of current M. truncatula genome.
Genome Mt4.0 and annotation Mt4.0v2 was used for computations

Figure 7: Chromosomal location of genes under putative positive selection in all M. truncatula populations, as determined by dN/dS analysis.

Figure 8: Predicted geographical location of the M. truncatula reference accession Jemalong-A17, using the plantGPS algorithm.
Closest accessions with reported geographical location are displayed in cyan. The predicted location is the centroid of the closest accessions, weighted by their genetic distance to A17.

|    | K1    | K2    | K3    | K4    | K5    | K6    | K7    |
|----|-------|-------|-------|-------|-------|-------|-------|
| K2 | 0.262 |       |       |       |       |       |       |
| K3 | 0.274 | 0.294 |       |       |       |       |       |
| K4 | 0.226 | 0.249 | 0.105 |       |       |       |       |
| K5 | 0.280 | 0.296 | 0.150 | 0.118 |       |       |       |
| K6 | 0.218 | 0.231 | 0.146 | 0.101 | 0.146 |       |       |
| K7 | 0.228 | 0.255 | 0.127 | 0.086 | 0.122 | 0.086 |       |
| K8 | 0.262 | 0.229 | 0.318 | 0.272 | 0.322 | 0.259 | 0.279 |

(a)

| Admixture component | $F_{IS}$ | Pop. size | Population name | Country |
|---|---|---|---|---|
| K1 | 0.53 | 11 | Algiers | Algeria |
| K2 | 0.69 | 23 | Spanish Coastal | Spain, Portugal |
| K3 | 0.54 | 15 | North Tunisian Coastal | Tunisia |
| K4 | 0.56 | 54 | Atlas | Algeria, Tunisia |
| K5 | 0.48 | 13 | South Tunisian Coastal | Tunisia |
| K6 | 0.62 | 29 | French | France |
| K7 | 0.58 | 63 | Greek | Greece and neighbours |
| K8 | 0.62 | 53 | Spanish-Morocco Inland | Spain, Morocco |

(b)

|    | Algiers | Spanish Coastal | N-Tunisian Coastal | Atlas | S-Tunisian Coastal | French | Greek |
|---|---|---|---|---|---|---|---|
| Spanish Coastal | 0.251 |  |  |  |  |  |  |
| N-Tunisian Coastal | 0.294 | 0.255 |  |  |  |  |  |
| Atlas | 0.228 | 0.218 | 0.060 |  |  |  |  |
| S-Tunisian Coastal | 0.322 | 0.275 | 0.112 | 0.090 |  |  |  |
| French | 0.204 | 0.176 | 0.102 | 0.070 | 0.116 |  |  |
| Greek | 0.236 | 0.220 | 0.084 | 0.060 | 0.082 | 0.048 |  |
| Spanish-Morocco Inland | 0.243 | 0.149 | 0.300 | 0.261 | 0.316 | 0.229 | 0.266 |

(c)

Table 1

|  | Estimate | Std. Error | t value | Pr(>|t|) |
|---:|---:|---:|---:|---:|
| Intercept | 2.4541 | 0.0861 | 28.50 | 0.0000 |
| Spanish Coastal | -1.4147 | 0.2290 | -6.18 | 0.0000 |
| South Tunisian Coastal | -0.7222 | 0.2367 | -3.05 | 0.0026 |
| Greek | 0.6017 | 0.1636 | 3.68 | 0.0003 |
| Spanish-Morocco Inland | -0.8518 | 0.1804 | -4.72 | 0.0000 |

Table 2

|  | diff | lwr | upr | p.adj |
|---|---|---|---|---|
| Reference R vs Reference S | 1.3286 | 0.9035 | 1.7537 | 0.0000 |
| Predicted R vs Predicted S | 1.2294 | 0.7070 | 1.7518 | 0.0000 |
| Reference R vs Predicted R | -0.2308 | -0.6859 | 0.2243 | 0.5544 |
| Reference S vs Predicted S | -0.1316 | -0.6280 | 0.3649 | 0.9018 |
| Reference S vs Predicted R | 1.0978 | 0.6211 | 1.5745 | 0.0000 |
| Reference R vs Predicted S | -1.4602 | -1.9360 | -0.9844 | 0.0000 |

Table 3

| Gene name | chrom | start | Pop.1 | | Pop.2 | | Pop.3 | | Pop.4 | | Pop.5 | | Pop.6 | | Pop.7 | | Pop.8 | |
|---|---|---|---|---|---|---|---|---|---|---|---|---|---|---|---|---|---|---|
| | | | dN/dS | pvalue | dN/dS | pvalue | dN/dS | pvalue | dN/dS | pvalue | dN/dS | pvalue | dN/dS | pvalue | dN/dS | pvalue | dN/dS | pvalue |
| Medtr1g045490 | chr1 | 17064668 | 0.25 | 3.5e-04 | 0.25 | 3.5e-04 | 0.25 | 3.5e-04 | 0.20 | 2.2e-05 | 0.25 | 3.5e-04 | 0.20 | 2.2e-05 | 0.20 | 2.2e-05 | 0.25 | 3.5e-04 |
| Medtr2g009110 | chr2 | 1780826 | 0.45 | 6.5e-08 | 0.45 | 6.5e-08 | 0.46 | 1.1e-07 | 0.45 | 6.5e-08 | 0.45 | 6.5e-08 | 0.45 | 6.5e-08 | 0.45 | 6.5e-08 | 0.45 | 6.5e-08 |
| Medtr3g460780 | chr3 | 23879139 | 0.12 | 1.1e-05 | 0.11 | 2.2e-06 | 0.12 | 1.1e-05 | 0.11 | 2.2e-06 | 0.12 | 1.1e-05 | 0.11 | 5.0e-06 | 0.11 | 2.2e-06 | 0.11 | 2.2e-06 |
| Medtr3g027940 | chr3 | 8822001 | 0.19 | 1.1e-05 | 0.19 | 1.1e-05 | 0.19 | 1.1e-05 | 0.19 | 1.1e-05 | 0.19 | 1.1e-05 | 0.19 | 1.1e-05 | 0.19 | 1.1e-05 | 0.19 | 1.1e-05 |
| Medtr4g078220 | chr4 | 30117318 | 0.07 | 1.1e-05 | 0.12 | 2.6e-05 | 0.13 | 5.7e-05 | 0.12 | 2.6e-05 | 0.13 | 5.7e-05 | 0.12 | 2.6e-05 | 0.12 | 2.6e-05 | 0.12 | 2.6e-05 |
| Medtr4g091580 | chr4 | 36293871 | 0.07 | 1.1e-05 | 0.07 | 1.1e-05 | 0.07 | 1.1e-05 | 0.07 | 1.1e-05 | 0.08 | 5.9e-05 | 0.07 | 2.5e-05 | 0.07 | 1.1e-05 | 0.07 | 2.5e-05 |
| Medtr5g037410 | chr5 | 16383754 | 0.29 | 6.3e-04 | 0.32 | 3.3e-04 | 0.17 | 2.2e-05 | 0.22 | 8.8e-05 | 0.18 | 8.4e-05 | 0.29 | 6.3e-04 | 0.35 | 1.2e-03 | 0.32 | 3.3e-04 |
| Medtr5g076220 | chr5 | 32463725 | 0.39 | 4.0e-04 | 0.33 | 2.5e-05 | 0.35 | 4.6e-05 | 0.33 | 2.5e-05 | 0.33 | 2.5e-05 | 0.33 | 2.5e-05 | 0.35 | 4.6e-05 | 0.33 | 2.5e-05 |
| Medtr6g033580 | chr6 | 10931616 | 0.06 | 1.9e-06 | 0.06 | 1.9e-06 | 0.06 | 1.9e-06 | 0.06 | 1.9e-06 | 0.06 | 4.5e-06 | 0.07 | 1.1e-05 | 0.06 | 1.9e-06 | 0.06 | 1.9e-06 |
| Medtr6g034870 | chr6 | 12162953 | 0.21 | 2.2e-08 | 0.21 | 2.2e-08 | 0.21 | 2.2e-08 | 0.21 | 2.2e-08 | 0.21 | 2.2e-08 | 0.21 | 2.2e-08 | 0.21 | 2.2e-08 | 0.21 | 2.2e-08 |
| Medtr6g034920 | chr6 | 12210408 | 0.13 | 5.7e-05 | 0.12 | 2.6e-05 | 0.12 | 2.6e-05 | 0.12 | 2.6e-05 | 0.13 | 5.7e-05 | 0.13 | 5.7e-05 | 0.12 | 2.6e-05 | 0.12 | 2.6e-05 |
| Medtr6g047750 | chr6 | 17217691 | 0.27 | 1.2e-06 | 0.28 | 1.1e-06 | 0.29 | 7.6e-06 | 0.28 | 1.1e-06 | 0.29 | 7.6e-06 | 0.28 | 1.1e-06 | 0.28 | 1.1e-06 | 0.28 | 1.1e-06 |
| Medtr6g011890 | chr6 | 3572293 | 0.37 | 2.0e-05 | 0.32 | 6.8e-06 | 0.36 | 3.7e-05 | 0.36 | 3.7e-05 | 0.36 | 3.7e-05 | 0.35 | 1.1e-05 | 0.34 | 2.1e-05 | 0.35 | 1.1e-05 |
| Medtr7g060720 | chr7 | 21955726 | 0.57 | 1.3e-04 | 0.57 | 1.3e-04 | 0.62 | 6.9e-04 | 0.57 | 1.3e-04 | 0.50 | 3.3e-05 | 0.57 | 1.3e-04 | 0.57 | 1.3e-04 | 0.57 | 1.3e-04 |
| Medtr7g023740 | chr7 | 7752286 | 0.14 | 1.3e-04 | 0.14 | 2.1e-06 | 0.14 | 1.3e-04 | 0.11 | 2.2e-06 | 0.11 | 2.2e-06 | 0.11 | 2.2e-06 | 0.14 | 9.5e-07 | 0.14 | 2.1e-06 |

Table 4